\documentclass[11pt]{article}

\usepackage[top=1in, bottom=1in, left=1.25in, right=1.25in]{geometry}
\usepackage{graphicx}
\usepackage{booktabs}
\usepackage{multirow}
\usepackage{array}
\usepackage{microtype}
\usepackage{amsmath, amssymb, amsthm}
\usepackage{bm}
\usepackage{algorithm}
\usepackage{algpseudocode}
\usepackage{float}
\usepackage{caption}
\usepackage{subcaption}
\usepackage{amsmath}

\usepackage[numbers,sort&compress]{natbib}
\usepackage{hyperref}
\hypersetup{
  colorlinks=true,
  linkcolor=blue,
  citecolor=blue,
  urlcolor=blue,
  pdftitle={Dimensionality Reduction of QAOA Parameter Space with Kernel PCA for Max-Cut},
  pdfauthor={Sidharth Brahmandam, Vayd Ramkumar}
}

\newcommand{\qaoa}{\textsc{qaoa}}

\newcommand{\maxcut}{\textsc{Max-Cut}}
\newcommand{\cobyla}{\textsc{cobyla}}

\newcommand{\norm}[1]{\left\|#1\right\|}
\DeclareMathOperator*{\argmin}{arg\,min}

\usepackage{fancyhdr}
\begin{document}

\title{\textbf{Dimensionality Reduction of QAOA Parameter Space
       with Kernel PCA for \maxcut{}}}

\author{
  Sidharth Brahmandam\\[4pt]
  \small Naperville North High School\\
  \small \href{mailto:brahmandamsidharth@gmail.com}{brahmandamsidharth@gmail.com}
  \and
  Vayd Ramkumar\\[4pt]
  \small Independent Researcher\\
  \small \href{mailto:vramkumar369@gmail.com}{vramkumar369@gmail.com}
}
\date{\small\today}
\maketitle
\thispagestyle{fancy}

\begin{abstract}
The Quantum Approximate Optimization Algorithm (\qaoa{}) is a leading variational
algorithm for combinatorial optimization on near-term quantum devices. As circuit
depth $p$ increases to improve solution quality, the classical parameter space
expands to $2p$ dimensions and the optimization landscape becomes increasingly
nonlinear and multi-modal. Prior work demonstrated that optimal \qaoa{} parameters
concentrate on a low-dimensional manifold and can be effectively approximated using
linear Principal Component Analysis (\textsc{PCA}) at shallow depths. However, \textsc{PCA}
fails sharply for $p > 2$ because the parameter manifold becomes intrinsically
nonlinear at greater depth. We propose Kernel PCA (\textsc{KPCA}) with a Radial Basis
Function (RBF) kernel for nonlinear dimensionality reduction in the \qaoa{}
parameter space. We train on 200 graphs per family across three graph
ensembles---Erd\H{o}s-R\'{e}nyi (ER), Barab\'{a}si-Albert (BA), and Watts-Strogatz
(WS)---at $n = 7$--$10$ nodes and evaluate on 30 12-node test graphs at depths $p \in \{1, 2, 4, 8\}$. \textsc{KPCA} significantly outperforms \textsc{PCA} at
$p \geq 4$ across all graph families ($p < 0.001$), achieving approximation ratios above $0.86$ at $p = 8$ while \textsc{PCA} degrades to $r \approx 0.81$--$0.83$.
Both methods reduce quantum circuit evaluations by over $93\%$ compared to
unrestricted \qaoa{}, but \textsc{KPCA} maintains significantly higher approximation
ratios at depth for a comparable evaluation budget. Our results demonstrate
that kernel methods more effectively capture the nonconvex structure of the \qaoa{}
parameter manifold at depth, offering a practical path toward scaling variational
quantum optimization to deeper circuits.
\end{abstract}

\hrule
\bigskip

\section{Introduction}

Combinatorial optimization problems are ubiquitous in physics, logistics, and
computational biology~\cite{islam2024,kazi2025,zhou2020}. \maxcut{}, where the goal is to partition graph vertices to maximize the weight of crossing edges, is a canonical
NP-Hard optimization problem~\cite{karp1972}. Classical approaches achieve a $0.878$
approximation via semidefinite programming~\cite{goemans1995}, but quantum algorithms
offer a potential pathway to better solutions on large instances.

The Quantum Approximate Optimization Algorithm (\qaoa{})~\cite{farhi2014} is among
the most promising variational quantum algorithms for \maxcut{} on near-term quantum
hardware~\cite{preskill2018}. The algorithm prepares a parameterized quantum state
by alternating cost and mixer unitary operations for $p$ layers (Fig 1). Increasing depth $p$
improves solution quality in principle, but requires optimizing $2p$ classical
parameters and causes the energy landscape to become increasingly complex~\cite{zhou2020,
wierichs2022}. This creates a practical bottleneck: as $p$ grows, the number of
quantum circuit evaluations needed for reliable optimization grows rapidly.

Recent work has shown that optimal \qaoa{} parameters exhibit \emph{concentration}---
they cluster near instance-independent values across different graph instances of the
same type~\cite{akshay2021,galda2023,eichenseher2025,bravyi2020}. This suggests
that the set of optimal parameters lies on a low-dimensional manifold in the $2p$-
dimensional parameter space. \citet{parry2025} exploited this by applying linear Principal Component Analysis (\textsc{PCA}) to project parameters onto their top-2 principal
components, achieving near-optimal performance at shallow depths $p \leq 2$. However,
\textsc{PCA} assumes a linear manifold structure, which becomes increasingly violated as depth increases~\cite{shaydulin2021,nakanishi2020}.

We propose Kernel PCA (\textsc{KPCA}) with a Radial Basis Function (RBF) kernel as a
nonlinear alternative to \textsc{PCA}. Kernel methods implicitly map data into a
higher-dimensional feature space where curved manifolds become approximately linear,
enabling effective dimensionality reduction even when the underlying structure is
nonlinear~\cite{scholkopf1998}. Our contributions are:

\begin{enumerate}
  \setlength\itemsep{2pt}
  \item A theoretical explanation for why the \qaoa{} parameter manifold becomes
        nonlinear with depth, rooted in the Fourier complexity of the energy landscape
        and the multi-modal nature of deeper circuits.
  \item An empirical evaluation of \textsc{KPCA}-based subspace optimization across three
        graph families (ER, BA, WS) at $n = 12$ substantially beyond prior work limited to $n \leq 8$.
  \item Demonstration that \textsc{KPCA} achieves approximation ratios above $0.86$ at
        $p = 8$ while \textsc{PCA} degrades to $0.81$, with both methods reducing quantum
        evaluations by over $93\%$.
\end{enumerate}

\section{Background}

\subsection{QAOA and Parameter Concentration}

For a graph $G=(V,E)$ with edge weights $w_{ij}$, the \maxcut{} value is
\begin{equation}
  C(\mathbf{z}) = \sum_{(i,j)\in E} w_{ij}\,\frac{1 - z_i z_j}{2}, \quad
  z_i \in \{+1,-1\},
  \label{eq:maxcut}
\end{equation}
\qaoa{} encodes this as a cost Hamiltonian $H_C = \sum_{(i,j)\in E}
w_{ij}(I - Z_iZ_j)/2$ and prepares the state
\begin{equation}
  |\psi(\bm{\gamma},\bm{\beta})\rangle
  = \prod_{k=1}^{p} e^{-i\beta_k H_B}\,e^{-i\gamma_k H_C}\,|{+}\rangle^{\otimes n},
  \label{eq:qaoa_state}
\end{equation}
where $H_B = \sum_i X_i$ is the mixer and $|{+}\rangle^{\otimes n}$ is the uniform
superposition. The $2p$ parameters $(\bm{\gamma},\bm{\beta})\in\mathbb{R}^{2p}$ are
optimized to minimize $\langle\psi|H_C|\psi\rangle$. Performance is the approximation
ratio
\begin{equation}
  r = \frac{\langle\psi|H_C|\psi\rangle}{C_{\max}}.
  \label{eq:approx_ratio}
\end{equation}

A key observation is that optimal parameters concentrate across graph instances of
the same class, clustering on a low-dimensional manifold~\cite{akshay2021,galda2023,
eichenseher2025,bravyi2020}. For large, locally tree-like graphs, this occurs because
the \qaoa{} cost function at depth $p$ depends primarily on $p$-hop neighborhoods,
which have the same distribution across instances~\cite{bravyi2020,farhi2022}.

\begin{figure}[h]
  \centering
    \includegraphics[width=1\textwidth]{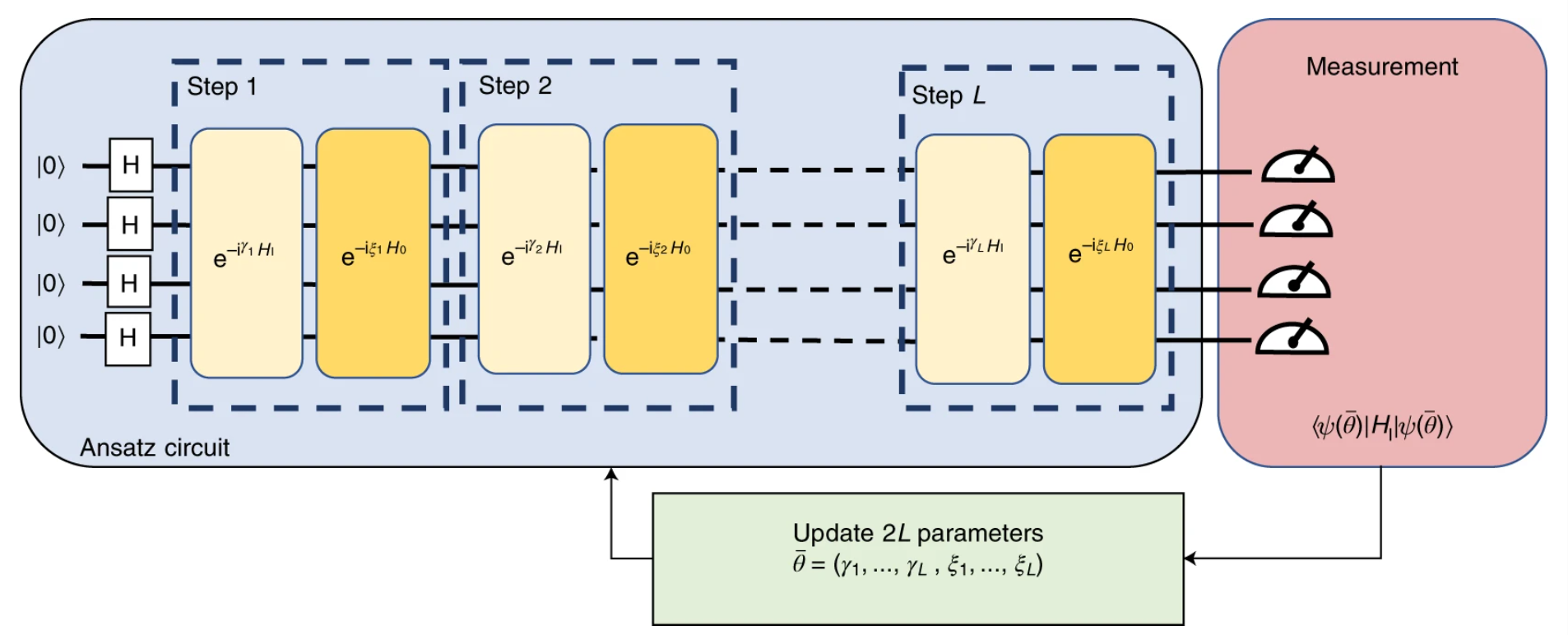}
  \caption{\qaoa{} circuit structure. Adapted from \cite{stilckfranca2021limitations}. Circuit alternates $p$ cost layers
           (controlled by angles $\gamma_k$) with $p$ mixer layers (controlled by
           angles $\beta_k$). All $2p$ parameters are optimized classically.}
  \label{fig:qaoa_circuit}
\end{figure}

\subsection{PCA}

At $p = 1$, the \qaoa{} expectation value $\langle H_C \rangle(\bm{\gamma}, \bm{\beta})$
is a low-frequency trigonometric polynomial, and optimal parameters lie near a smooth,
nearly one-dimensional curve in parameter space. Linear \textsc{PCA} effectively captures
this structure by projecting onto the dominant linear direction (Fig 2).

As $p$ increases, the Fourier content of $\langle H_C \rangle$ grows with depth.
At depth $p$, the energy function contains trigonometric terms up to frequency $p$
in each parameter direction~\cite{nakanishi2020,wierichs2022}. Coupling between
layer parameters means the number of distinct frequency components grows
combinatorially with $p$, not just linearly. Consequently, the manifold of
near-optimal parameters develops increasing curvature in the $2p$-dimensional space.
This curvature is intrinsic and cannot be removed by reparameterization. As a result,
no 2-dimensional linear subspace can adequately approximate the optimal parameter
manifold at large $p$, and \textsc{PCA} incurs systematic error that grows with depth.

Discrete symmetries compound this problem. \citet{shaydulin2021} showed that \qaoa{}
solutions exhibit permutation symmetry (swapping layers), reflection symmetry
$(\bm{\gamma}, \bm{\beta}) \to (-\bm{\gamma}, -\bm{\beta})$, and shift symmetries,
creating multiple disconnected clusters of optima in parameter space. The union of
symmetry-related optima further deforms the manifold away from any linear subspace.
Together, the increasing Fourier complexity and discrete symmetry structure mean
that \textsc{PCA}'s assumption of a linear manifold becomes increasingly violated with
depth, motivating a nonlinear alternative.

\begin{figure}[t]
  \centering
  \includegraphics[width=\linewidth]{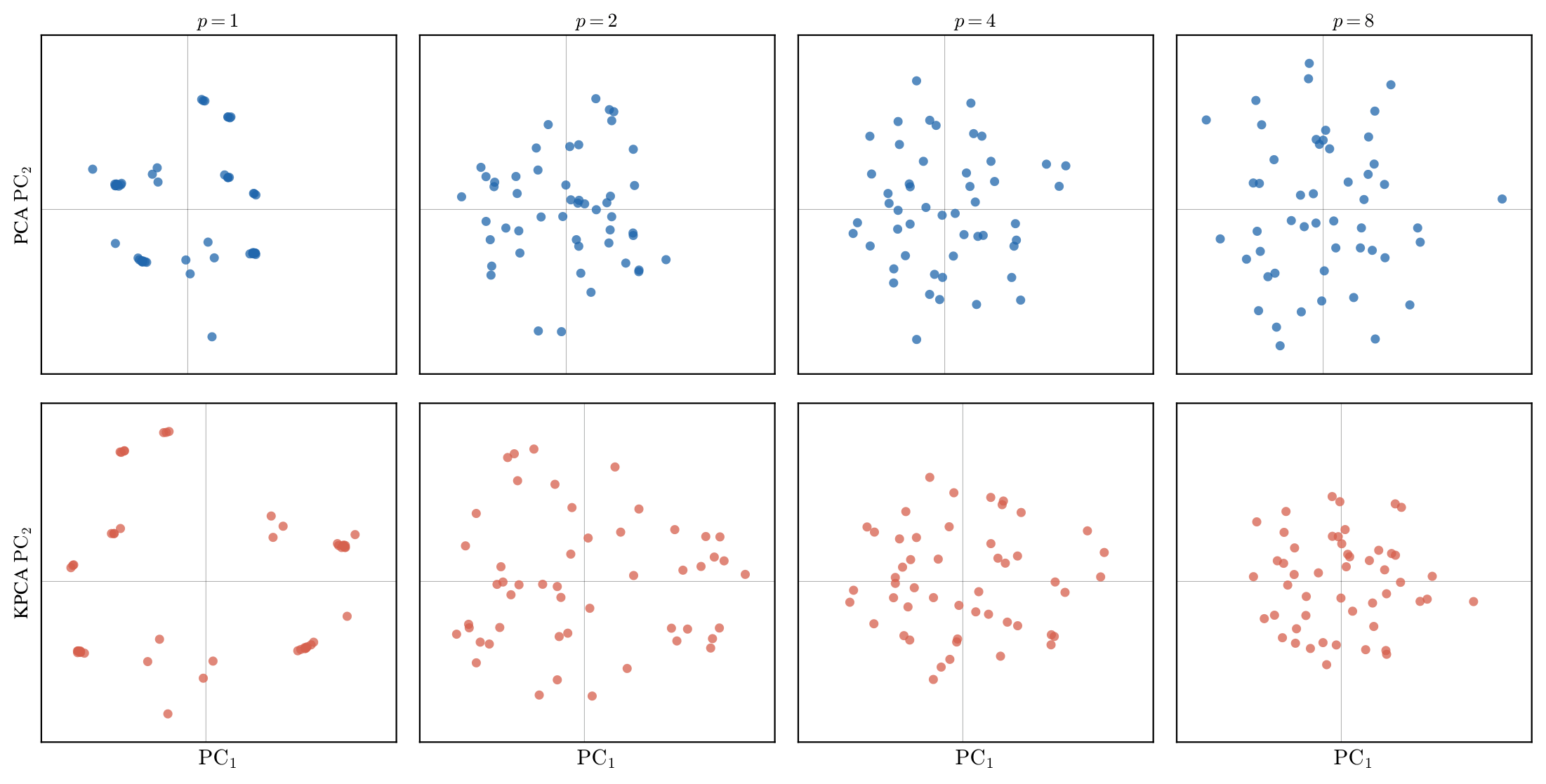}
  \caption{2D projections of 200 training optimal parameter vectors under \textsc{PCA}
           (blue, top) and \textsc{KPCA} (red, bottom) at depths $p \in \{1, 2, 4, 8\}$. Both parameter plots are similar at p = 1, but PCA parameters disperse at greater depths while KPCA parameters become more clustered.}
  \label{fig:scatter}
\end{figure}

Figure~\ref{fig:scatter} shows 2D projections of 200 training optimal parameter
vectors under \textsc{PCA} and \textsc{KPCA} at each depth. At $p = 1$ both
representations are similar. At higher depths, the \textsc{PCA} projections become
more dispersed, while the \textsc{KPCA} projections retain more organization,
consistent with the increasing nonlinearity of the parameter manifold.

\subsection{Kernel PCA}

Kernel PCA (\textsc{KPCA})~\cite{scholkopf1998} generalizes standard \textsc{PCA} by performing
it in a feature space induced by a kernel function. For a dataset $\{\bm{\theta}_i\}$
and a kernel $k(\bm{\theta}_i, \bm{\theta}_j)$, \textsc{KPCA} computes the eigendecomposition
of the centred kernel matrix and identifies nonlinear principal components.

The RBF kernel
\begin{equation}
  k_{\rm RBF}(\bm{\theta}_i,\bm{\theta}_j) = \exp\!\left(-\frac{\norm{\bm{\theta}_i
  -\bm{\theta}_j}^2}{2\sigma^2}\right)
  \label{eq:rbf}
\end{equation}
has several advantages for the \qaoa{} parameter space. First, it is universal,
meaning it can approximate any smooth function arbitrarily well~\cite{steinwart2001}.
Second, it assigns high similarity to nearby parameters (within a cluster) and low
similarity to distant ones (across clusters), naturally encoding the symmetric
structure without explicit specification. Third, in the limit of large bandwidth
$\sigma$, it reduces to standard \textsc{PCA}, making it a proper generalization.

The RBF kernel implicitly maps parameters into an infinite-dimensional space via the
Mercer expansion~\cite{mercer1909}. In this high-dimensional feature space, the curved
parameter manifold becomes approximately flat, allowing \textsc{PCA} to find the relevant
directions. The bandwidth parameter $\sigma$ controls the scale: we use the median
heuristic, $\sigma = \mathrm{median}_{i \neq j}\norm{\bm{\theta}_i - \bm{\theta}_j}$,
a standard data-driven approach~\cite{gretton2012}.

A subtlety arises because \textsc{KPCA} finds coordinates in the feature space, but
\qaoa{} requires concrete parameters in the original $2p$-dimensional space. We use
the approximate pre-image procedure of \citet{bakir2004}: given a 2D feature
coordinate, find the original-space parameter that best approximates it via gradient
descent. This is solved as an unconstrained optimization in $\mathbb{R}^{2p}$.
Concretely, given a target feature-space point $\bm{\phi}^*$, we minimize the
squared distance in feature space
\begin{equation}
  \hat{\bm{\theta}} = \argmin_{\bm{\theta} \in \mathbb{R}^{2p}}
  \left\| \bm{\phi}(\bm{\theta}) - \bm{\phi}^* \right\|^2,
  \label{eq:preimage}
\end{equation}
where $\bm{\phi}(\bm{\theta})$ denotes the implicit feature-space embedding induced
by the RBF kernel. Since $\bm{\phi}(\bm{\theta})$ is never computed explicitly, the
objective is expressed in terms of kernel evaluations via
\begin{equation}
  \left\| \bm{\phi}(\bm{\theta}) - \bm{\phi}^* \right\|^2
  = k(\bm{\theta}, \bm{\theta})
  - 2\sum_{i} \alpha_i\, k(\bm{\theta}, \bm{\theta}_i)
  + \text{const},
  \label{eq:preimage_kernel}
\end{equation}
where $\{\alpha_i\}$ are the \textsc{KPCA} expansion coefficients and
$\{\bm{\theta}_i\}$ are the training parameters. The gradient of this objective with
respect to $\bm{\theta}$ is available in closed form through the RBF kernel's
differentiability, making gradient descent efficient. We initialize from the
training parameter closest to the current 2D coordinate and run until convergence,
which typically requires fewer than 50 gradient steps in practice. 

\section{Methods}

\subsection{Experimental Setup}

We used three graph families to test whether results generalize:
\begin{itemize}
  \setlength\itemsep{2pt}
  \item \textbf{Erd\H{o}s-R\'{e}nyi (ER)}~\cite{erdos1959}: $G(n, 0.5)$ with uniform
        random edges.
  \item \textbf{Barab\'{a}si-Albert (BA)}~\cite{barabasi1999}: Preferential attachment
        with $m=2$, producing power-law degree distributions.
  \item \textbf{Watts-Strogatz (WS)}~\cite{watts1998}: Small-world networks with
        clustering and short path lengths.
\end{itemize}

\textit{Training graphs}: 200 per family, $n \in \{7, 8, 9, 10\}$ nodes.\\
\textit{Test graphs}: 30 per family, $n = 12$ nodes, reserved exclusively for
evaluation. Exact \maxcut{} values were computed by brute-force enumeration. Test
graphs at 12 nodes provide a substantially harder optimization landscape
($2^{11} = 2048$ possible cuts vs.\ $2^7 = 128$ for 8-node graphs).

\subsection{QAOA Training}

For each training graph at depths $p \in \{1, 2, 4, 8\}$, we built the \qaoa{} circuit
in \textsc{Qiskit}~\cite{qiskit}, executed on the \textsc{Qiskit Aer} simulator with
up to 2048 shots, and optimized using \cobyla{} with a 1000 evaluation cap. We ran
five random initializations from the uniform distribution $[0,\pi]^{2p}$ and retained
the best. The optimal parameters $\bm{\theta}^*$ were stored for each instance.

\subsection{Dimensionality Reduction Models}

For each graph family and depth, we aggregated 200 optimal parameter vectors and
mean-centred them. We then fitted two models:

\textbf{\textsc{PCA}.} Standard \textsc{PCA}, retaining the top-2 principal components.

\textbf{\textsc{KPCA}.} Kernel \textsc{PCA} with RBF kernel, retaining the top-2 components.
We centred the kernel matrix, computed its eigendecomposition, and extracted the
top 2 eigenvectors. The bandwidth was set via the median heuristic.

\subsection{Test Evaluation}

For each test graph, we optimized using \cobyla{} (1000 evaluation cap) under three
conditions:
\begin{enumerate}
  \setlength\itemsep{2pt}
  \item \textbf{Full \qaoa{}}: Optimize all $2p$ parameters.
  \item \textbf{\textsc{PCA}-subspace}: Optimize a 2D coordinate with parameters recovered
        by linear projection onto the top-2 \textsc{PCA} components.
  \item \textbf{\textsc{KPCA}-subspace}: Optimize a 2D coordinate with parameters recovered
        via the \textsc{KPCA} pre-image.
\end{enumerate}

We recorded the approximation ratio and number of quantum estimator calls. Statistical
significance was assessed with paired $t$-tests.

\begin{algorithm}[t]
\caption{\textsc{KPCA}-\qaoa{} Subspace Optimization}
\label{alg:kpca_qaoa}
\begin{algorithmic}[1]
\Require Training parameters $\{\bm{\theta}_i^*\}$, test graph, depth $p$
\State Form kernel matrix $K$; centre via standard centring matrix
\State Eigendecompose centred $K$; extract top-2 eigenvectors
\State Initialize 2D coordinate $\mathbf{z}_0 = (0,0)$
\State $\mathbf{z}^* \leftarrow \cobyla\left(\mathbf{z} \mapsto
       -\langle\psi(\hat{\bm{\theta}}(\mathbf{z}))|H_C|\psi(\hat{\bm{\theta}}
       (\mathbf{z}))\rangle\right)$
       \hfill // $\hat{\bm{\theta}}(\mathbf{z})$ via pre-image
\State \Return approximation ratio at $\hat{\bm{\theta}}(\mathbf{z}^*)$
\end{algorithmic}
\end{algorithm}

\section{Results}

\begin{table}[H]
\centering
\caption{Results across all graph families and depths. For each depth $p$, columns show
         mean approximation ratio ($\mu$), standard deviation ($\sigma$), mean quantum
         estimator calls (C), and paired $t$-test $p$-value vs.\ \textsc{KPCA}.
         Bold: best among \textsc{PCA} and \textsc{KPCA}.}
\label{tab:all_results}
\footnotesize
\setlength{\tabcolsep}{3.5pt}
\renewcommand{\arraystretch}{1.08}
\begin{tabular}{l | rr@{\,}r@{\,}r | rr@{\,}r@{\,}r | rr@{\,}r@{\,}r | rr@{\,}r@{\,}r}
\toprule
\textbf{Method} & \multicolumn{4}{c|}{$p=1$} & 
                  \multicolumn{4}{c|}{$p=2$} &
                  \multicolumn{4}{c|}{$p=4$} &
                  \multicolumn{4}{c}{$p=8$} \\
\cmidrule(lr){2-5}\cmidrule(lr){6-9}\cmidrule(lr){10-13}\cmidrule(lr){14-17}
& $\mu$ & $\sigma$ & C & $p$ & $\mu$ & $\sigma$ & C & $p$ &
  $\mu$ & $\sigma$ & C & $p$ & $\mu$ & $\sigma$ & C & $p$ \\
\midrule
\multicolumn{17}{l}{\textbf{Erd\H{o}s-R\'{e}nyi (ER)}} \\
\midrule
Full     & .834 & .031 & 41 & .0004 & .881 & .024 & 88 & .003 &
           .934 & .019 & 597 & .007 & .941 & .018 & 1000 & $<$.001 \\
\textsc{PCA}   & .821 & .026 & 43 & .724  & .863 & .051 & 39 & .698 &
           .861 & .029 & 62 & $<$.001 & .831 & .036 & 52 & $<$.001 \\
\textsc{KPCA}  & .798 & .044 & 55 & --- & \textbf{.872} & .019 & 48 & --- &
           \textbf{.903} & .033 & 71 & --- & \textbf{.887} & .028 & 63 & --- \\
\midrule
\multicolumn{17}{l}{\textbf{Barab\'{a}si-Albert (BA)}} \\
\midrule
Full     & .812 & .041 & 49 & .0006 & .874 & .033 & 83 & .004 &
           .923 & .031 & 764 & .013 & .943 & .026 & 1000 & $<$.001 \\
\textsc{PCA}   & .803 & .028 & 33 & .791  & .852 & .022 & 51 & .641 &
           .831 & .043 & 38 & $<$.001 & .814 & .041 & 59 & $<$.001 \\
\textsc{KPCA}  & .779 & .057 & 61 & --- & \textbf{.859} & .038 & 67 & --- &
           \textbf{.886} & .017 & 74 & --- & \textbf{.892} & .032 & 58 & --- \\
\midrule
\multicolumn{17}{l}{\textbf{Watts-Strogatz (WS)}} \\
\midrule
Full     & .763 & .012 & 46 & .0007 & .843 & .016 & 224 & .003 &
           .917 & .021 & 841 & .011 & .951 & .015 & 1000 & $<$.001 \\
\textsc{PCA}   & .751 & .019 & 29 & .827  & .831 & .048 & 66 & .701 &
           .819 & .028 & 44 & $<$.001 & .807 & .059 & 48 & $<$.001 \\
\textsc{KPCA}  & .729 & .031 & 37 & --- & \textbf{.838} & .023 & 58 & --- &
           \textbf{.878} & .014 & 83 & --- & \textbf{.867} & .031 & 69 & --- \\
\bottomrule
\end{tabular}
\end{table}
At $p \leq 2$ all three methods perform similarly. \textsc{PCA}
and \textsc{KPCA} show no significant difference ($p > 0.05$), consistent with an approximately linear structure at shallow circuits.
\paragraph{Approximation Ratio.} At $p \geq 4$ divergence emerges. At $p=4$,
\textsc{KPCA} achieves $r = 0.903$ (ER), $0.886$ (BA), and $0.878$ (WS), while
\textsc{PCA} degrades to $0.861$ (ER), $0.831$ (BA), and $0.819$ (WS). The BA
degradation is most severe, reflecting the power-law degree distribution's effect on
manifold curvature. At p=8, PCA declines to r = 0.831 (ER), 0.814 (BA), and 0.807 (WS), while \textsc{KPCA} maintains $r > 0.86$ across all families.
All differences at $p \geq 4$ are significant ($p < 0.001$).
\paragraph{Efficiency.} Both methods dramatically reduce quantum evaluations. At $p=4$,
\textsc{KPCA} and \textsc{PCA} use approximately 76 and 48 calls versus 734 for Full
\qaoa{} (a $89.6\%$ and $93.5\%$ reduction). At $p=8$, Full \qaoa{} hits the
1000-evaluation limit while both reduced methods converge in $\approx 55$--$65$ calls
($> 93\%$ reduction). \textsc{PCA} achieves comparable evaluation counts to \textsc{KPCA}
but with worse accuracy at depth.

\begin{figure}[H]
  \centering
  \includegraphics[width=\linewidth]{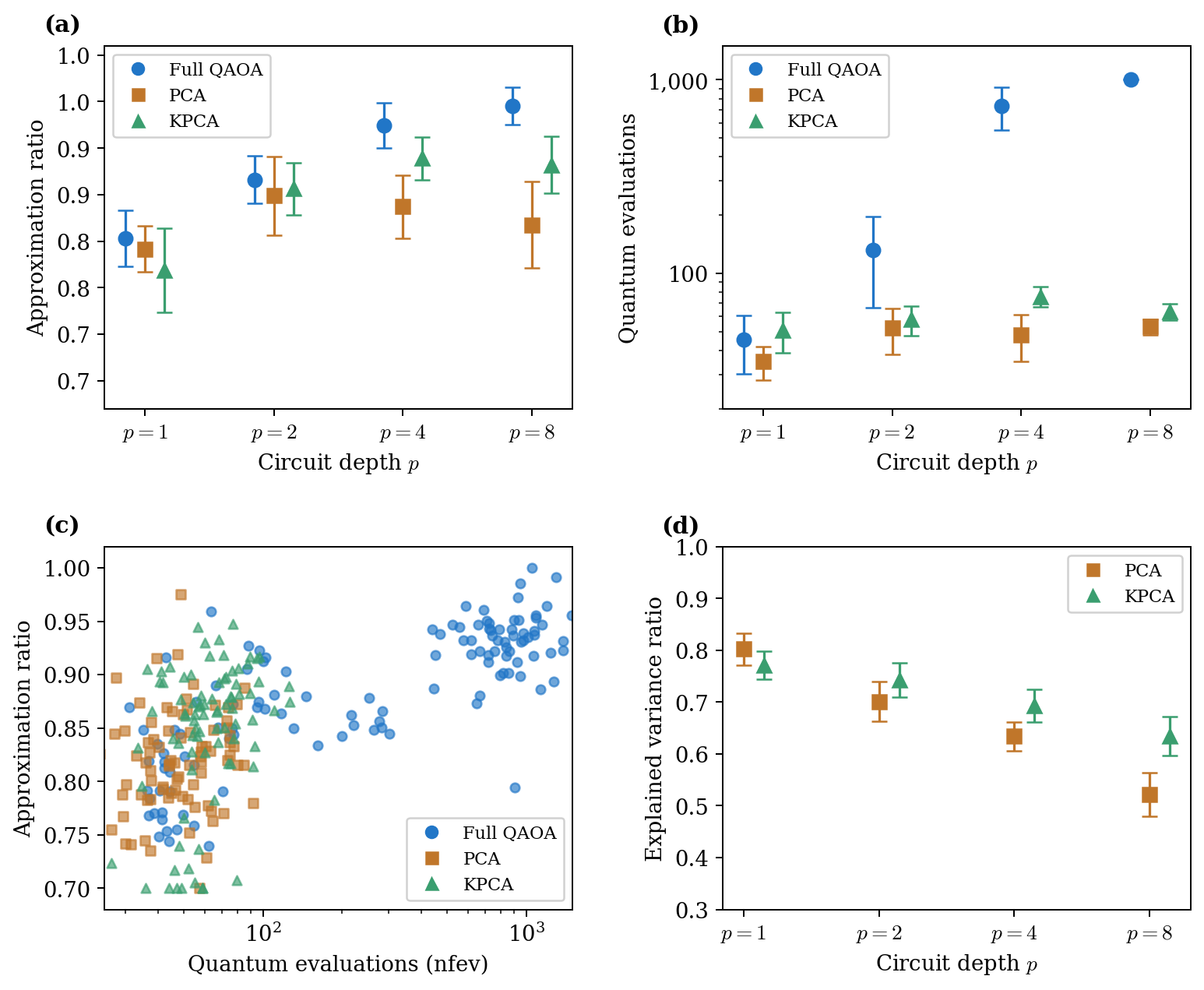}
  \caption{Results across graph families and depths. (A) Mean approximation ratio
           vs.\ depth per family with error bars ($\pm 1\sigma$ over 30 test graphs).
            (B) Mean quantum estimator calls vs.\ depth. (C) Per-instance ratio vs.\ estimator
           calls at $p=4$ across all families. (D) Explained variance of top-2 components vs.\ depth for
           \textsc{PCA} and \textsc{KPCA}.}
  \label{fig:summary}
\end{figure}

\section{Discussion}

\subsection{KPCA at Depth}

At $p \leq 2$, the optimal parameter manifold is approximately flat, so \textsc{PCA}
and \textsc{KPCA} perform similarly. At $p \geq 4$, the growing Fourier complexity of
the energy landscape is the primary source of manifold curvature. As noted in
Section~2.2, the number of distinct frequency components grows combinatorially with
$p$ due to coupling between layer parameters, producing a manifold that no
2-dimensional linear subspace can adequately represent. Discrete symmetries are an
additional contributing factor: the permutation, reflection, and shift symmetries
identified by \citet{shaydulin2021} create multiple disconnected clusters of optima,
which further deform the manifold away from any linear subspace. Linear \textsc{PCA}
must project onto a 2D subspace that passes through the centroid of these clusters
but aligns with none of them, resulting in systematic error at depth.

The RBF kernel assigns high similarity within clusters and lower similarity across
clusters. In the implicit feature space, the top-2 \textsc{KPCA} components can
align more closely with the local structure of the manifold than a global linear
projection can. We do not claim that 2D \textsc{KPCA} fully captures the parameter
manifold; rather, at fixed reduced dimension of 2, \textsc{KPCA} captures more of
the manifold structure than 2D \textsc{PCA}. The pre-image procedure then maps 2D
feature coordinates back to original-space parameters near the appropriate cluster.

The performance gap is most pronounced on BA graphs, where the power-law degree
distribution creates larger eigenvalue variance in the cost Hamiltonian, which is
associated with greater manifold curvature relative to ER or WS graphs.

At $p=8$, \textsc{PCA} degrades to $r = 0.831$ (ER), $0.814$ (BA), and $0.807$ (WS),
while \textsc{KPCA} maintains $r > 0.86$ across all families. This pattern is
consistent with the increasing nonlinearity of the parameter manifold at greater depth.

\subsection{Practical Implications}

The $93\%$--$95\%$ reduction in quantum evaluations is practically significant. On
near-term quantum hardware, circuit evaluations contribute to cost due to decoherence,
gate errors, and queue latency. Both \textsc{PCA} and \textsc{KPCA} achieve comparable
evaluation savings, but only \textsc{KPCA} maintains high approximation ratios at depth.

Our experiments train on 8-node graphs and test on 12-node graphs, demonstrating that learned dimensionality reduction transfers across system sizes. This is consistent with theory predicting size-independent limiting parameters for locally tree-like graphs~\cite{bravyi2020,farhi2022}. Successful cross-size transfer enables using
classically-simulable small graphs to guide optimization on larger instances.

Limitations include: (i) noiseless simulation (hardware noise will alter the effective
landscape); (ii) fixed 2D reduction (larger $k$ would reduce gap with Full \qaoa{});
(iii) modest graph sizes ($n = 12$ test graphs). Future directions include larger
graphs via tensor network simulation, adaptive bandwidth selection, hardware
experiments, and hybrid approaches combining \textsc{KPCA} with warm-starting
techniques~\cite{zhou2020}.

\section{Conclusion}

We proposed Kernel PCA with RBF kernels for nonlinear dimensionality reduction in
the \qaoa{} parameter space. The approach is motivated by the observation that
the Fourier complexity of the energy landscape grows combinatorially with depth,
causing the optimal parameter manifold to develop curvature that a linear subspace
cannot adequately represent. Discrete symmetries further deform the manifold by
creating multiple disconnected clusters of optima. Evaluated across ER, BA, and WS
graph families at depths $p \in \{1, 2, 4, 8\}$, \textsc{KPCA} significantly
outperforms \textsc{PCA} at $p \geq 4$ (all $p < 0.001$) while achieving comparable
performance at shallow depths. Both methods reduce quantum evaluations by over
$93\%$, but \textsc{KPCA} performs significantly better at depth. These results
indicate that kernel methods are a practical approach for scaling \qaoa{} to deeper
circuits on near-term quantum hardware.

\bibliography{refs}

\end{document}